\documentclass[a4paper,twosides]{article}
\usepackage{CJK,multicol,multirow,graphics,fancyhdr,epstopdf}
\usepackage{amsmath,amsfonts,amssymb,bm,upgreek,mathrsfs,mathcomp}
\usepackage[pagewise,switch,columnwise]{lineno}
\usepackage[compress,nospace]{cite}
\usepackage[dvipsnames]{xcolor}
\usepackage{CPL-2022}
\usepackage[dvips]{graphicx}
\RequirePackage{lineno}

\newcommand{\cplyear}{2023} \newcommand{\cplvol}{40}
 \newcommand{\cplpagenumber}{121301}
 \newcommand{\cplpage}{\cplpagenumber-\thepage}

\usepackage{caption}
\usepackage[font=small]{caption}
\usepackage{indentfirst}

\begin{document}

\begin{CJK}{GBK}{song}\vspace* {-4mm}
\begin{center}
\large\bf{\boldmath{Status and prospects of exotic hadrons at Belle II}}
\footnotetext{\hspace*{-5.4mm}$^{*}$Corresponding authors. Email: shencp@fudan.edu.cn

\noindent\copyright\,{\cplyear}
\href{http://www.cps-net.org.cn}{Chinese Physical Society} and
\href{http://www.iop.org}{IOP Publishing Ltd}}
\\[5mm]
\normalsize \rm{}Sen Jia$^{1}$, Weitao Xiong$^{2}$, and Chengping Shen$^{2,3*}$
\\[3mm]\small\sl $^{1}$School of Physics, Southeast University, Nanjing 211189, China

$^{2}$Key Laboratory of Nuclear Physics and Ion-beam Application (MOE) and Institute of Modern Physics, Fudan University, Shanghai 200443, China

$^{3}$School of Physics, Henan Normal University, Xinxiang 453007, China

\normalsize\rm{}(Received xxx; accepted manuscript online xxx)
\end{center}
\end{CJK}
\vskip 1.5mm

\small{\narrower In the past twenty years, many new hadrons that are difficult to be explained within the conventional quark model have been discovered in the quarkonium region, which are called exotic hadrons. Belle II experiment, as the next-generation $B$ factory, provides a good platform to explore them. The charmonium-like states  can be produced at Belle II in several ways, such as $B$ meson decays, initial-state radiation processes, two-photon collisions, and double charmonium productions.
The bottomonium-like states can be produced directly in $e^+e^-$ colliding energies at Belle II with low continuum backgrounds. Belle II plans to perform a high-statistics energy scan from the $B\bar B$ threshold up to the highest possible energy of 11.24 GeV to search for new $Y_b$ states with $J^{PC}$ = $1^{--}$, $X_b$ (the bottom counterpart of $\chi_{c1}(3872)$ (also known as $X(3872)$)), and partners of $Z_b$ states. In this paper, we give a mini-review on the
status and prospects of exotic hadrons at Belle II.

\par}\vskip 3mm
\noindent{\narrower{DOI: 10.1088/0256-307X/40/12/121301}

\par}\vskip 5mm

\section{Introduction}

Hadron is a composite subatomic particle made of two or more quarks held together by the strong interaction, which can be described by Quantum Chromodynamics (QCD) in the Standard Model.
At high energies, the asymptotic freedom guarantees that properties of QCD can be computed in a weak-coupling perturbation theory\ucite{1343,1346}.~While, at low energies, the weak-coupling perturbation theory fails to interpret the color confinement\ucite{1}.
Therefore, the study of hadrons in charmonium and bottomonium regions offers many opportunities for insight on the nonperturbative behavior of QCD.

Besides the conventional mesons composed with $q\bar q$ and baryons with $qqq$, multiquark states with $qq\bar q\bar q$ or $qqqq\bar q$ have been proposed by Gell-Mann in 1964 in one of the first publications on the quark model\ucite{214}.
In this paper, we refer to any state which does not coincide with the expectations for an ordinary $q\bar q$ or $qqq$ hadron in the quark model as `exotic'. In the charmonium and bottomonium regions, the exotic hadrons are also called charmonium-like and bottomonium-like states.

There is not progress in searching for exotic hadrons until the observation of $X(3872)$ in the $\pi^+\pi^-J/\psi$ final states by Belle in 2003\ucite{262001}. 
Here, the $X(3872)$ denotes the $\chi_{c1}(3872)$ in particle data group (PDG)\ucite{PDG}.
We used the $X(3872)$ instead of $\chi_{c1}(3872)$ thereinafter.
The properties of the $X(3872)$ did not fit those of an ordinary charmonium.
Since then, many new exotic hadrons have been discovered by various experiments\ucite{1,026201}.
Among them, Belle experiment contributed a lot.
Besides the $X(3872)$, the first charged charmonium-like state, $Z_c(4430)^{\pm}\to\pi^{\pm}\psi(2S)$\ucite{142001}, was observed by Belle in $B$ decays in 2007, which was confirmed by LHCb seven years later\ucite{222002}.
Almost at the same time, the $Z_c(3900)^{\pm}\to \pi^{\pm} J/\psi$ was discovered by BESIII in direct $e^+e^-$ annihilations and Belle in the
initial state radiation (ISR) process, which is regarded as an alternative tetraquark state\ucite{252001,252002}.
Up to now, only two bottomonium-like states, $Z_b(10610)^{\pm}$ and $Z_b(10650)^{\pm}$, were discovered\ucite{122001}. They were first found in $\pi^{\pm}\Upsilon(nS)$ $(n=1,2,3)$ final states by Belle using $\Upsilon(5S)$ data samples.

Belle II experiment has started to collect ``physics" data since 2019.
In June 2022, the accelerator SuperKEKB achieved the highest instantaneous luminosity
of 4.7$\times 10^{34}$ cm$^{-2}$s$^{-1}$ in the world.
The data collected at Belle II by the end of 2023 corresponds to an integrated luminosity of 427.79 fb$^{-1}$.
The integrated luminosity is expected to be increased drastically in the following years, and will ultimately reach 50 ab$^{-1}$ around 2034.
With large data samples, the experimental studies on exotic hadrons at Belle II should proceed:
searches for new exotic hadrons and expected partners of existing exotic hadrons, searches for new decay
channels of known states, and detailed measurements of all accessible properties, including
spin-parities, absolute branching fractions, line-shapes, etc.
We do not discuss the exotic baryons due to the limited condition at Belle II.

\section{Achievements at Belle}

The Belle experiment ran at the KEKB $e^+e^-$ asymmetric energy collider at the KEK Laboratory, Tsukuba, Japan between 1999 to 2010\ucite{479,499,2013}.
A large part of data samples at Belle were recorded on or near the $\Upsilon(4S)$ resonance, where copious $B\bar B$ pairs  used to study the $B$ physics are produced.
In addition, Belle collected a series of special data sets at $\Upsilon(1S)$, $\Upsilon(2S)$, and $\Upsilon(10860)$ resonances.
The data samples recorded at Belle are detailed in Table~\ref{tab1}\ucite{2012}. The total integrated luminosity reaches approximately 1 ab$^{-1}$.
All of the data sets were used to explore the hadron spectroscopy, and a large number of
new conventional quarkonia and some exotic hadrons were found using these samples.

\begin{table*}[htbp!]
\caption{Integrated luminosities of Belle data samples. The scan data were collected with approximately 1 fb$^{-1}$ per point in the $\Upsilon(10860)$ and $\Upsilon(11020)$ energy ranges from 10.63 to 11.02 GeV\ucite{2012}.}
\centering
\small
\label{tab1}
\begin{tabular}{c  c  c  c}
\hline\hline
\multirow{2}*{Resonance} & On-peak & Off-peak & \multirow{2}*{Number of accumulated events} \\
& luminosity (fb$^{-1}$) & luminosity (fb$^{-1}$) & \\\hline
$\Upsilon(1S)$ & 5.7 & 1.8 & $102\times10^6$ \\
$\Upsilon(2S)$ & 24.9 & 1.7 & $158\times10^6$ \\
$\Upsilon(3S)$ & 2.9 & 0.25 & $11\times10^6$ \\
$\Upsilon(4S)$ & 711.0 & 89.4 & $772\times10^6$ $B\bar B$ \\
$\Upsilon(5S)$ & 121.4 & 1.7 & $7\times10^6$ $B_s\bar B_s$ \\
Energy scan & - & 27.6 & \\
\hline\hline
\end{tabular}
\end{table*}

\begin{figure}[htbp]
\centering
\includegraphics[scale=0.7]{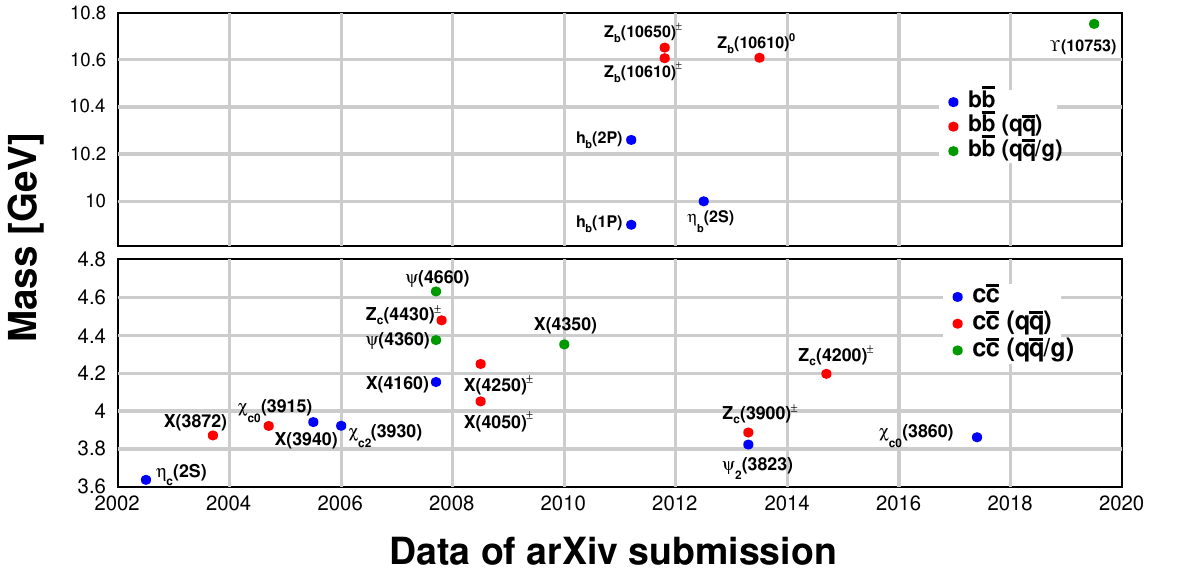}
\caption{Quarkonium(-like) states observed by Belle as a function of the year of observation\ucite{QWG}.}\label{fig1}
\end{figure}

Belle is constantly enriching the hadron spectroscopy. As shown in Fig.~\ref{fig1}\ucite{QWG}, many quarkonium(-like) states were observed at Belle, including long-term missing conventional quarkonia and new exotic hadrons of $X$, $Y$, $Z_c$, and $Z_b$ states.
In the following, we introduce three outstanding achievements on exotic hadrons at Belle.

In 2003, Belle observed a narrow charmonium-like state decaying into $\pi^+\pi^-J/\psi$ with a mass of $(3872.0\pm0.8)$ MeV/$c^2$ in $B^{\pm}\to K^{\pm}\pi^+\pi^-J/\psi$\ucite{262001}.
This is the first state with unconventional charmonium properties, thus opening a new era for studying of exotic hadrons.
The $X(3872)$ was subsequently confirmed by several other experiments\ucite{072001,162002,071103}.
The $J^{PC}$ = $1^{++}$ was determined by LHCb by performing a five-dimensional angular correlation analysis in $B$ decays\ucite{011102}.
So far $X(3872)$ is still one of the most interesting exotic meson candidates because of its unexpected properties as follows.
The mass of $X(3872)$ is close to the $D \bar D^{(*)}$ threshold, and its width is very narrow. Such property is very different from the excited $\psi$ states, such as $\psi(3770)$, $\psi(4040)$, and $\psi(4160)$, which decays into $D^{(*)}\bar D^{(*)}$ with large branching fractions, but their widths are very large.
The enhancement of isospin-violating $J/\psi\rho$ decay was observed.
Very recently, Belle determined that
the lower limit on the $D \bar D^*$ coupling constant is 0.094 at 90\% credibility level, which implies the partial width of $D \bar D^*$ in $X(3872)$ decays is much larger than those of $J/\psi\rho$, $J/\psi\omega$, and radiative decay modes\ucite{112011}.

Searching for charged charmonium-like states is one of the most promising ways in studying exotic hadrons, since
such a state must contain at least four quarks and thus cannot be a conventional quark-antiquark meson.
The first charged charmonium-like state, $Z_c(4430)^{\pm}$, was reported in the $\pi^{\pm}\psi(2S)$ mass spectrum in $B\to K\pi^{\pm}\psi(2S)$
decays at Belle in 2007\ucite{142001}. It was not confirmed by LHCb until seven years later\ucite{222002}.
Unlike $Z_c(4430)^{\pm}$, $Z_c(3900)$ state decaying into $\pi^{\pm}J/\psi$ was observed simultaneously
by BESIII and Belle\ucite{252001,252002}. The $Z_c(4200)^+\to\pi^+J/\psi$ with $J^P = 1^+$ was observed by performing an amplitude analysis of $\bar B^0\to J/\psi K^-\pi^+$ decays in four dimensions by Belle\ucite{112009}, which has been confirmed by LHCb\ucite{152002}.

In comparison with the number of charmonium-like states from PDG\ucite{PDG}, the number of bottomonium-like states is smaller due to
lower production rates and limited experimental conditions. Up to now, only two bottomonium-like states were observed: $Z_b(10610)$ and $Z_b(10650)$.~They were discovered by Belle only using $\Upsilon(5S)$ on-resonance and scan data samples\ucite{122001,072003,Zb142001,212001,052016}.
The $Z_b(10610)^+$ and $Z_b(10650)^+$ can decay into $\pi^+\Upsilon(nS)$ ($n$ = 1, 2, 3),  $\pi^+h_b(mP)$ ($m$ = 1, 2), and $B^{(*)+}\bar B^{*0}$\ucite{122001,072003,Zb142001,212001}.
The decay modes of $Z_b(10610)^+\to B^{+}\bar B^{*0}$ and $Z_b(10650)^+\to B^{*+}\bar B^{*0}$ are dominant\ucite{212001}.
The neutral partner of the $Z_b(10610)^+$ was observed as well in its $\pi^0\Upsilon(2S)$ and $\pi^0\Upsilon(3S)$ decay modes\ucite{052016}.

\section{Belle II experiment and data taking}

Belle II, the first super $B$-factory experiment, is operating at the asymmetric-energy $e^+e^-$ SuperKEKB collider\ucite{188}.
Two beams with energies of 7.0 GeV and 4.0 GeV collide at an angle of 83 mrad (22 mrad in KEKB).
It presents several Belle subdetectors in an upgraded version together with other ones newly developed for Belle II\ucite{1011.0352}. The vertex detector consists of pixel sensors and double-sided silicon vertex detectors. The central drift chamber is surrounded by two types of Cherenkov light counters: an azimuthal array of time-of-propagation detector for the barrel region and an aerogel ring-imaging Cherenkov detector for the forward endcap region to the electron beam. The Belle electromagnetic CsI(Tl) crystal calorimeter, with upgraded electronics, is reused in Belle II along with the solenoid and the iron flux return yoke.
The muon and $K^0_L$ identification system with upgraded glass-electrode resistive plate chambers and scintillator is the outermost and only sub-detector outside the magnet.

\begin{figure}[htbp]
\centering
\includegraphics[width=8cm]{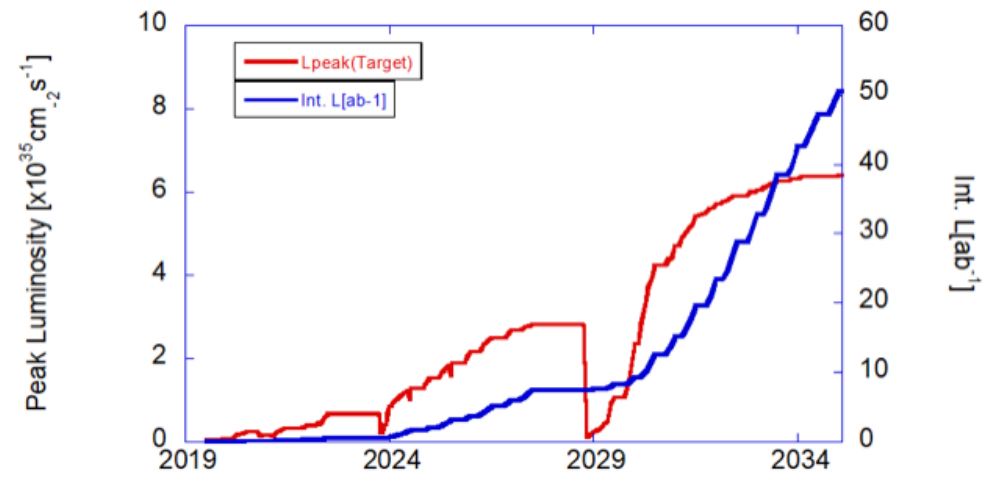}
\caption{Data taking plan at Belle II\ucite{2207.06307}. The red line shows the instantaneous luminosity, and the blue line shows the integrated luminosity.}\label{Lumi}
\end{figure}

Taking a ``nano-beam" scheme of reduction in the beam size and increase in the currents, the luminosity at SuperKEKB is expected to increase drastically\ucite{188}.
Figure~\ref{Lumi} shows the data taking plan at Belle II\ucite{2207.06307}.
Belle II's goal is to accumulate 50 ab$^{-1}$ of data until 2035\ucite{123C01,2207.06307}. This schedule includes two long shutdowns (LS1
and LS2) for detector upgrades. During LS1, Belle II will complete the installation of the second layer of
the pixel-based inner vertex detector. For LS2, major upgrades to the detector are being considered, along
with extensive upgrades to the accelerator.
After 2023, the size of data collected at Belle II will increase sharply with the completion of LS1.
Although the data will be recorded at $\Upsilon(4S)$ resonance mainly, data taking at other energy points with substantial
statistics can be guaranteed. It is worth mentioning that all of the data at Belle II can be used to probe exotic hadrons discussed below.

\section{Prospects for charmonium-like states at Belle II}

There are four ways to explore the charmonium-like states at Belle II: $B$ meson decays, ISR processes, two-photon collisions, and double charmonium productions.
Charmonium(-like) states ($X_{c\bar c}$) are produced in $B$ meson decays in association with a kaon: $B\to K X_{c\bar c}$, which is a Cabibbo-Kobayashi-Maskawa favoured process with a relatively large branching fraction of $10^{-4}$ -- $10^{-3}$\ucite{PDG}.
Using an ISR technique, the charmonium-like states with $J^{PC}=1^{--}$ below $e^+e^-$ center-of-mass (C.M.) energy can be explored and a continuous mass spectrum in a wide region can be studied\ucite{181}.
The two-photon interaction at Belle II via $e^+e^-\to e^+e^-\gamma^*\gamma^*\to e^+e^- X_{c\bar c}$ is a typical and effective tool to study the charmonium-like states with various $J^{P}$ quantum numbers.
The double charmonium productions, such as $e^+e^-\to J/\psi X_{c\bar c}$\ucite{cc142001}, provides a good platform for understanding the perturbative and non-perturbative effects in QCD.
In the subsections below, we show four promising measurements related to charmonium-like states at Belle II in the near future.

\subsection{The absolute branching fraction of $X(3872)$ decay}

The measurement of absolute branching fraction of $X(3872)$ decay can provide key information to identify the nature of $X(3872)$.
The original tetraquark model predicted the branching fraction of $X(3872) \to \pi^+ \pi^- J/\psi$ to be about 50\%\ucite{014028}.
By contrast, various molecular models predicted this branching fraction to be $\leq$10\%\ucite{054022,054008,124107,091401,021301}.

Measurement of absolute branching fractions for $X(3872)$ via $B\to K X(3872)$ is unique at $B$ factories.
By exclusively reconstructing one $B$ meson ($B_{\rm tag}$) and identifying the $K$ in the decay of the other $B$ meson, the missing mass technique can be used to identify $X(3872)$. Then one can obtain the branching fraction for $B\to KX(3872)$, and further calculate the absolute branching fractions of $X(3872)$ decaying into different final states. No significant $X(3872)$ signal was observed in the recoil mass distribution of $K$ at Belle\ucite{012005}.
Evidence is found by the BaBar Collaboration for the decay $B\to K X(3872)$ at a $3\sigma$ level by allowing all $B$ combinations in an event and performing two layers of neural networks to suppress backgrounds\ucite{152001}. The branching fraction of $B^+\to K^+ X(3872)$ was measured to be $(2.1\pm0.8)\times 10^{-4}$ for the first time. In addition, the branching fraction of $X(3872)\to\pi^+\pi^-J/\psi$ was calculated to be $(4.1\pm1.3)\%$.

Two improvements are achieved at Belle II: (1) an upgraded Full Event Interpretation (an exclusive reconstruction algorithm)\ucite{6} is used to reconstruct thousands of final states for $B_{\rm tag}$ candidates to increase the efficiency; (2) a unique multiple variable analysis of DeepSets based on the neural network\ucite{1703.06114} is applied to suppress continuum backgrounds and reject the secondary kaons produced in $B$-daughter $D$ meson decays. The above method may reject about 95\% of backgrounds in the $X(3872)$ signal region (81\% at BaBar\ucite{152001}). We expect about 600 $X(3872)$ candidates from the  $B\to KX(3872)$ decays per 1 ab$^{-1}$ can be observed at Belle II. Therefore, measurements of absolute branching fractions of $X(3872)$ at Belle II with much larger data samples in the near future are promising. Surely, the measurements of absolute branching fractions for other charmonium(-like) states are also feasible at Belle II.

\subsection{Search for new $Y$ states in $e^+e^-\to K^+K^-J/\psi$ via ISR}

\begin{figure}[htbp]
\centering
\includegraphics[scale=0.4]{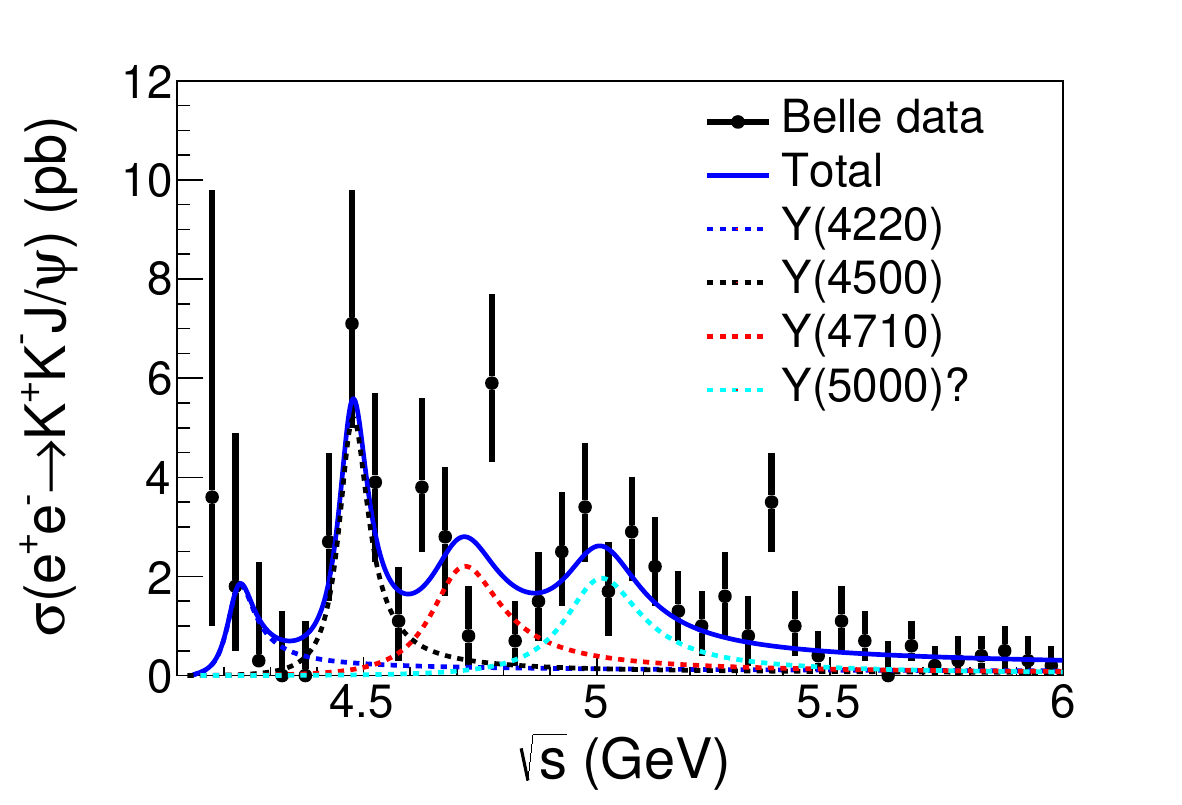}
\caption{The measured $e^+e^-\to K^+K^-J/\psi$ cross sections by Belle (dots with error bars)\ucite{072015}. The $Y(4220)$, $Y(4500)$, $Y(4710)$, and $Y(5000)$ components are indicated with dashed curves.}\label{KKjpsi}
\end{figure}

The $Y$ states with specific quantum number $J^{PC}$ = $1^{--}$ always attract much attention among charmonium-like states.
The cross sections for $e^+e^-\to K^+K^-J/\psi$ ($\sigma(e^+e^-\to K^+K^-J/\psi)$) were measured via ISR using the full Belle data, as shown by the dots with error bars in Fig.~\ref{KKjpsi}\ucite{072015}. From Fig.~\ref{KKjpsi}, there seems to be abundant structures, although the statistics is limited.
Recently, BESIII studied the cross sections for $e^+e^-\to K^+K^-J/\psi$ and $e^+e^-\to K^0_SK^0_SJ/\psi$, and found the $Y(4220)$, $Y(4500)$, and $Y(4710)$ states\ucite{092005,111002}.
To describe the energy dependency of $\sigma(e^+e^-\to K^+K^-J/\psi)$ in Fig.~\ref{KKjpsi}, we include the $Y(4220)$, $Y(4500)$, and $Y(4710)$ components with their masses and widths fixed according to Refs.~\cite{092005,111002}, as shown by the blue, black, and red dashed curves in Fig.~\ref{KKjpsi}.
In addition, an extra resonance around 5 GeV, called the $Y(5000)$, needs to be included to coincide the higher side of the $\sigma(e^+e^-\to K^+K^-J/\psi)$ distribution. At Belle II with much larger data samples, one should measure the $Y(4220),~Y(4500),~Y(4710)$ decaying into $K^+K^-J/\psi$ with higher precision and verify whether there are new resonances with higher masses.

\begin{figure}[htbp]
\centering
\includegraphics[scale=0.4]{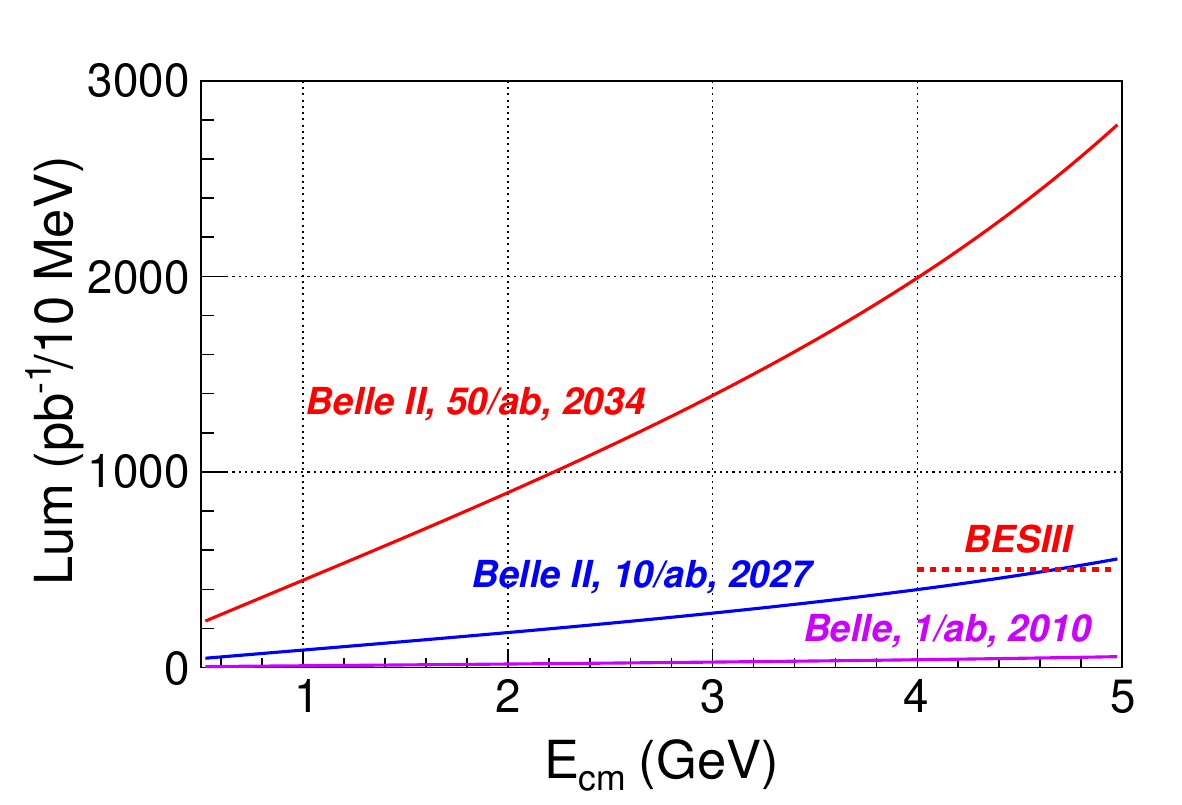}
\caption{The effective integrated luminosities at BESIII, Belle, and Belle II.}\label{ISRlum}
\end{figure}

In the ISR process, the effective integrated luminosity is decreased with the C.M.\ energy decreasing, as indicated in Fig.~\ref{ISRlum}.~The detection efficiency is also smaller compared to that in direct $e^+e^-$ annihilations at BESIII. However, Belle II can provide a continuous mass spectrum instead of a point-by-point one, which helps extract the mass and width for the charmonium-like state precisely. At Belle II, one can focus on the higher mass region to search for possible $Y$ states decaying into $K\bar D^{(*)}_sD^{(*)}$, $\bar D^{(*)}_sD_{s1}(2460)$, $\bar D^{(*)}_sD_{s1}(2536)$, $\bar D^{(*)}_sD^*_{s2}(2573)$, $\bar D^{*}_sD^*_{s0}(2317)$, $\bar \Lambda_c\Lambda_c$, $\bar \Lambda_c\Sigma_c$, $\bar\Sigma_c\Sigma_c$, $\bar \Xi_c\Xi_c$, etc. The intermediate states $Z_{cs}\to KJ/\psi$ and $Z_{cs}\to \bar D^{(*)}_sD^{(*)}$ can be also searched for in above processes.

\subsection{Confirmation of $X(4350)$ and determination of $J^P$ for $X(3915)$ in two-photon collisions}

The $X(4140)\to\phi J/\psi$ was first reported by CDF in $B^+\to K^+\phi J/\psi$\ucite{242002,1750139}.
Then, Belle and BaBar searched for such state in the same $B$ decay, but did not confirm its presence\ucite{615,012003}.
CMS, D0, and LHCb also searched for the $X(4140)$, and found its signal significance larger than $3\sigma$ (D0\ucite{012004}) or $5\sigma$ (CMS and LHCb\ucite{261,022003}).
In the full amplitude analysis from LHCb, the $X(4140)$, $X(4274)$, $X(4500)$, and $X(4700)$ were observed\ucite{022003}.
The structures in the $\phi J/\psi$ mass spectrum seem very rich. Therefore, it is worth studying the $\phi J/\psi$ mass spectrum in other processes.

The $\gamma\gamma\to\phi J/\psi$ process was measured between the threshold to 5 GeV/$c^2$ using a data sample of 825 fb$^{-1}$ at Belle\ucite{112004}. A narrow peak was observed around 4.530 GeV, named as $X(4350)$.
The signal yield is $8.8^{+4.2}_{-3.2}$ with a significance of 3.2$\sigma$ including systematic uncertainty.
This state has not been yet confirmed by other experiments.
It should be noted that the productions of the $X(4140)$ and $X(4274)$ with $J^{P}$ = $1^+$ observed by LHCb\ucite{022003} in two-photon fusions are forbidden according to the Landau-Yang theorem.
When the data collected at Belle II reaches 1~ab$^{-1}$, together with all of the Belle data, the significance of $X(4350)$ is expected to exceed $5\sigma$ as an observation.

The $X(3915)$ was originally seen by Belle in its $\omega J/\psi$ decay mode and was produced in both $B$ decays and two-photon collisions\ucite{182002,092001}.
The mass and width of $X(3915)$ in $\gamma\gamma\to\omega J/\psi$ were measured to be $(3915\pm4)$ MeV/$c^2$ and $(17\pm11)$ MeV\ucite{092001}.
Subsequently, BaBar confirmed its existence  and supported the assignment $J^{P}$ = $0^{+}$ based on a spin-parity analysis, thus identifying $X(3915)$ as a $\chi_{c0}(2P)$ state\ucite{072002}. However, the properties of $X(3915)$ do not fit expectations for the $\chi_{c0}(2P)$ since it has a mass 190 MeV/$c^2$ above the $S$-wave $D\bar D$ threshold but only has a width of 20 MeV\ucite{091501,057501}.
Moreover, in Ref.~\cite{022001}, the authors proposed that the spin-parity of $X(3915)$ should be $2^{++}$, if the $X(3915)$ has a non-$c\bar c$ admixture.
Recently, LHCb performed an amplitude analysis of the $B^+\to D^+D^-K^+$ decay and found a new spin-0 charmonium resonance around 3.92 GeV/$c^2$\ucite{112003}. If it is the same state as observed in the $\omega J/\psi$ decay, $X(3915)$ can be regarded as a $\chi_{c0}(2P)$ state.
To figure out this long-term unsolved issue on the $J^P$ of $X(3915)$, the angular distribution in $\gamma\gamma\to \omega J/\psi$ at Belle II with high precision can provide effective experimental information.

\section{Prospects for bottomonium-like states at Belle II}

There are at least four ways to explore the bottomonium-like states at Belle II: direct $e^+e^-$ annihilations, ISR processes, hadronic transitions, and radiative transitions. In $e^+e^-$ collisions, the C.M.\ energy can be set at some specific energy points to precisely measure the masses and widths of known resonances, and search for possible new states.~The $\Upsilon(10753)$ was discovered using the direct $e^+e^-$ collision data and ISR process from 10.52 to 11.02 GeV at Belle\ucite{220}. The hadronic and radiative transitions are effective ways to search for new bottomonium-like states.
The $h_b(1P)$ and $h_b(2P)$ were observed in the $\pi^+\pi^-$ transitions from $\Upsilon(10860)$ at Belle\ucite{Zb142001,032001}.
The $Z_c(10610)$ and $Z_c(10650)$ were observed in the pion transitions from $\Upsilon(10860)$ and $\Upsilon(11020)$ at Belle\ucite{122001,072003,Zb142001,052016}.
The $\eta_b(2S)$ was evident via the radiative decay of $h_b(2P)$ in $e^+e^- \to\Upsilon(5S)\to h_b(2P)(\to \gamma\eta_b(2S))\pi^+\pi^-$ at Belle\ucite{232002}.

\subsection{The study of $\Upsilon(10753)$}

The $\Upsilon(10753)$ was first observed by Belle in the energy dependence of cross sections for $e^+e^-\to\pi^+\pi^-\Upsilon(nS)$ ($n$ = 1, 2, 3) with a significance of $5.2\sigma$ including systematic uncertainties\ucite{220}. The mass and width of $\Upsilon(10753)$ are $(10752.7^{+5.9}_{-6.0})$ MeV/$c^2$ and $(35.5^{+18.0}_{-11.8})$ MeV.
Its existence is further supported by a fit to the `dressed' cross sections $\sigma(e^+e^-\to b\bar b)$ at C.M.\ energies $\sqrt{s}$ from 10.6 to 11.2 GeV\ucite{083001}.
The $\Upsilon(10753)$ was interpreted as a conventional bottomonium\ucite{074007,034036,59,014020,014036,357,04049,135340,11915,103845}, hybrid\ucite{034019,1}, or tetraquark state\ucite{135217,074507,11475,123102,381}. The nature of $\Upsilon(10753)$ has not been identified yet.

\begin{figure}[htbp]
\centering
\includegraphics[scale=0.5]{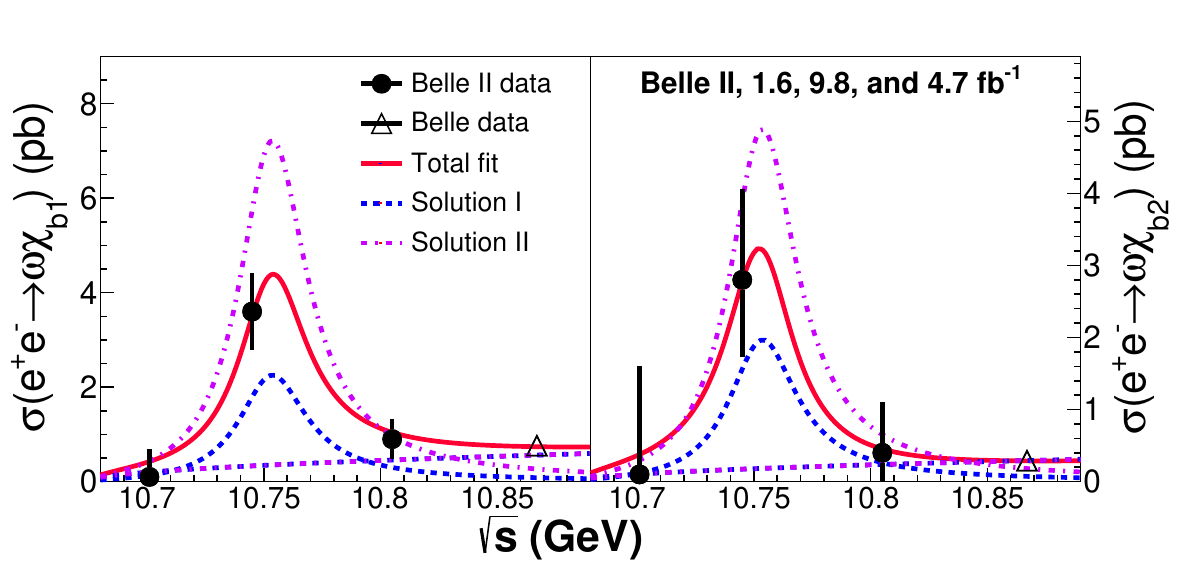}
\caption{The Born cross sections of (left) $e^+e^-\to\omega\chi_{b1}(1P)$ and (right) $e^+e^-\to\omega\chi_{b2}(1P)$  as a function of $\sqrt{s}$\ucite{091902}. The red solid curves show the total fits. The dashed curves show two solution results.}\label{omegachib}
\end{figure}

To further understand the nature of $\Upsilon(10753)$, Belle II collected 19 fb$^{-1}$ of unique data at four energy points around 10.75 GeV in November 2021. Using these data, the measurement of $e^+e^-\to\omega\chi_{bJ}(1P)$ ($J$ = 0, 1, 2) was performed\ucite{091902}.~The Born cross sections of $e^+e^-\to\omega\chi_{b1,b2}(1P)$ as a function of $\sqrt{s}$ are shown in Fig.~\ref{omegachib}, where the $\Upsilon(10753)\to \omega\chi_{b1,b2}(1P)$ signal is clear. The Born cross section ratio between $e^+e^-\to\omega\chi_{b1,b2}(1P)$ and $e^+e^-\to\pi^+\pi^-\Upsilon(nS)$ ($n$ = 1, 2, 3) is about 1.5 at $\sqrt{s}$ = 10.745 GeV and 0.15 at $\sqrt{s}$ = 10.867 GeV\ucite{091902,chib142001}. This indicates different internal structures for the $\Upsilon(10753)$ and $\Upsilon(10860)$, which otherwise have the same quantum numbers and are only 110 MeV/$c^2$ apart.

Using above data samples, $e^+e^-\to B\bar B$, $B\bar B^*$, and $B^*\bar B^*$ signals at $\sqrt{s}$ = 10.653, 10.701, 10.745, and 10.805 GeV were observed\ucite{slides}. By combining the Belle results at $\sqrt{s}$ from 10.63 to 11.02 GeV\ucite{137} and Belle II results at above four energy points\ucite{slides}, the energy dependencies of cross sections for $e^+e^-\to B\bar B$, $B\bar B^*$, and $B^*\bar B^*$ were obtained.
A new and unexpected observation is that the $e^+e^-\to B^*\bar B^*$ cross section increases rapidly just above the threshold, which can be explained by presence of a resonance or a bound state of $B^*\bar B^*$ near the $B^*\bar B^*$ threshold\ucite{2779,012002}.
Using above data samples, $\eta_b(1S)$ and $\chi_{b0}(1P)$ were searched for in the $\omega$ recoil mass distribution. No significant signals were observed, and upper limits at 90\% credibility level on the Born cross sections were set to be 2.5 pb and 8.6 pb for $e^+e^-\to \omega\eta_b(1S)$ and $e^+e^-\to \omega\chi_{b0}(1P)$ at $\sqrt{s}$ = 10.745 GeV, respectively\ucite{slides}.

More analyses are ongoing, including the searches for $e^+e^-\to\gamma X_b$ ($X_b$ is the bottom partner of $X(3872)$), $\Upsilon(10753)\to \eta\Upsilon(nS)$ ($n$ = 1, 2, 3), $\Upsilon(10753)\to K^+K^-\Upsilon(1S)$, and $\Upsilon(10753)\to\gamma\chi_{bJ}(1P)$, etc.
These measurements help us understand the internal structure of $\Upsilon(10753)$, and enrich our knowledge of the bottomonium spectroscopy.

\subsection{Bottomonium-like states above $B\bar B$ mass threshold}

Below $B\bar B$ threshold, bottomonia are described well by the potential models in QCD\ucite{1534,2981,074027}.
Above $B\bar B$ threshold, bottomonia demonstrate unexpected properties as follows.
There is no position for $Z_b(10610)$ and $Z_b(10650)$ in bottomonium potential model.
In comparison with the charmonium-like states, the number of observed bottomonium-like states is much smaller\ucite{PDG}.
The decay $\Upsilon(4S)\to\eta h_b(1P)$, forbidden in Heavy Quark Effective Theory\ucite{19,455}, has a larger probability than that of $\pi^+\pi^-\Upsilon(nS)$ ($n$ = 1, 2, 3)\ucite{U142001}.
For $\Upsilon(5S)$ and $\Upsilon(6S)$, their widths to $\pi^+\pi^-\Upsilon(nS)$ and $K^+K^-\Upsilon(nS)$ are several orders of magnitude larger than those of $\Upsilon(4S)$\ucite{112001}.
At this point, there is not enough experimental information for a systematic solution to this problem in the case of bottomonia.
The relevant measurements with data collected in bottomonium energy region at Belle II will provide additional information.

At Belle II, an energy scan from the $B\bar B$ threshold up to the highest possible energy of 11.24 GeV with $\sim$10 fb$^{-1}$ per point
and $\sim$10 MeV step is of high interest to search for new excited $\Upsilon$ and $Y_b$ (bottomonium-like states with $J^{PC}$ = $1^{--}$) states\ucite{123C01}. If the new bottomonium(-like) states are observed, about 500 fb$^{-1}$ data will be collected at corresponding energy point to perform a detailed study of radiative and hadronic transitions, to search for $X_b$ and other $Z_b$ states.
Unlike the $X(3872)$ decays, due to negligible isospin-breaking effects, $X_b$ may decay preferentially into $\pi^+\pi^-\pi^0\Upsilon(nS)$, $\pi^+\pi^-\chi_{bJ}(1P)$, and $\gamma\Upsilon(nS)$, instead of $\pi^+\pi^-\Upsilon(1S)$\ucite{1600,3063,034005,115122001}.
Although open flavour channels dominate the $Z_b$ decays, one can still search for more $Z_b$ states in $\pi\Upsilon(nS)$, $\pi h_b(nP)$, and $\rho\eta_b(nS)$ systems due to a higher reconstruction efficiency and free continuum contribution\ucite{7,1610.01102}.
Combined coupled-channel analyses of all exclusive cross sections will allow to determine the pole positions of the $\Upsilon(nS)$ and $Y_b$ states, their electronic widths, and couplings to various channels. Increase of maximal energy of 11.24 GeV by at least 100 MeV will allow to explore the
$\Lambda_b\bar \Lambda_b$ threshold and to search for baryon-antibaryon molecular states\ucite{123C01}. In the $\Lambda_b$ decays, the exotic baryons $P_c$ could be investigated at Belle II.

\section{Summary}

Belle has made great achievements in exotic hadrons.
The experimental studies at Belle II will proceed along the same lines as at Belle.
Measurements of absolute branching fractions for $X_{c\bar c}$ via $B\to K X_{c\bar c}$ are unique at Belle II with an advanced multiple variable analysis based on the neural network.~Besides $B$ decays, ISR technique, two-photon collisions, and double charmonium productions are also effective tools to explore the charmonium-like states. Belle II will collect abundant data around $\Upsilon(10860)$ and $\Upsilon(11020)$ resonances, which provide an ideal platform to study the bottomonium-like states above $B\bar B$ threshold. The $Y_b$, $X_b$, and $Z_b$ states can be widely investigated using these data in the near future.

\vspace{0.5cm}
{\it Acknowledgments. ---} This work is supported by National Natural Science Foundation of China (NSFC) under contract Nos.~12342024, 12135005, 12135005, 11975076, 12005040; and the Fundamental Research Funds for the Central Universities Grant RF.~1028623046.

\end{document}